\documentclass[5p,times,twocolumn]{elsarticle}

\usepackage{amsmath,amssymb}
\usepackage{graphicx}
\usepackage{microtype}
\usepackage{balance}
\usepackage{xcolor}
\usepackage[colorlinks=true,allcolors=blue]{hyperref}
\biboptions{numbers,sort&compress}
\newcommand{\GeV}{\,\mathrm{GeV}}

\journal{arXiv:2604.25542}

\begin{document}

\begin{frontmatter}

\title{A Finite-History Interpretation of the AMS-02 Positron Spectrum}

\author[as,ncku]{Yi Yang\corref{cor1}}
\cortext[cor1]{Corresponding author}
\ead{yiyang429@as.edu.tw}

\affiliation[as]{organization={Institute of Physics, Academia Sinica},
                 city={Taipei},
                 postcode={11529},
                 country={Taiwan}}
\affiliation[ncku]{organization={Department of Physics, National Cheng Kung University},
                   city={Tainan},
                   postcode={70101},
                   country={Taiwan}}

\begin{abstract}
I examine whether the separation between the characteristic energy scales of cosmic-ray electrons and positrons can be understood as a finite-history effect, without introducing a separate dominant source specifically to generate the high-energy positron feature. In this description the positron retains the opposite Dirac phase orientation relative to ordinary matter clocks, while local interactions and positive physical energies remain unchanged. Reduced accumulated overlap with the matter-defined Galactic environment is represented by a single, approximately shape-preserving energy rescaling. An illustrative overlap benchmark moves the broad structure of an empirical electron reference near 10 GeV into the few-hundred-GeV region. A number-conserving spatial-dilution example provides an order-of-magnitude interpretation of the relative amplitude. The comparison uses the published AMS-02 electron spectrum directly, with fixed, rounded horizontal and vertical scales rather than optimized spectral parameters. Known populations, including pulsars, may contribute subleading components in this interpretation. The resulting characteristic-scale displacement and geometrical amplitude interpretation provide a possible physical account of the positron spectral hierarchy.
\end{abstract}

\begin{keyword}
antiparticles \sep Dirac phase history \sep cosmic rays \sep positrons \sep AMS-02
\end{keyword}
\end{frontmatter}

\section{Introduction}
\label{sec:intro}

The AMS-02 electron and positron spectra have different energy dependences. The positron spectrum departs from its low-energy trend above about 25 GeV and develops a dropoff in the few-hundred-GeV region, whereas the electron spectrum has a different high-energy structure \cite{AMSpositron,AMSelectron}. The question addressed here is whether this spectral hierarchy can be understood through a different accumulated positron response, without adding a separate dominant source specifically to produce the high-energy feature.

The starting point is the opposite coordinate-phase orientation of particle and antiparticle modes in the Dirac field \cite{Stueckelberg,Feynman,Peskin}. The time-orientation formulation of Ref.~\cite{YangTime} retains this information relative to ordinary matter clocks. Applied to a finite Galactic environment, the picture explored here assigns the opposite phase history a reduced effective overlap with that environment. The microscopic interactions remain the same; the distinction concerns the accumulated history. The term ``advanced'' in the title refers to this opposite phase-history orientation, not to a trajectory running backward in laboratory time. A related use of finite-history geometry in a cosmological setting was discussed in Ref.~\cite{YangYeung}.

For the corresponding electron and positron component, I take a common underlying injection spectrum, including equal initial normalization in the geometric illustration below. This is the charge-symmetric component being compared: the observed difference is attributed to the subsequent response. Ordinary primary electrons and contributions from known source populations remain part of the physical setting. In particular, pulsars can supply energetic pairs \cite{HooperPulsars}. The present interpretation treats source-specific contributions as subleading over the high-energy interval illustrated, rather than requiring a specially introduced dominant positron source. Their fractions are not determined here.

The main consequence is simple: rescaling the energy axis of a spectrum with a feature moves that feature. The aim is a qualitative account of the lepton spectral scales and relative sizes, using rounded benchmarks and direct data comparisons rather than optimized spectral parameters or a detailed Galactic propagation calculation. This self-contained, positron-focused presentation shares its lepton analysis and Fig.~\ref{fig:positron} with the combined study of Ref.~\cite{YangCombined}, which also develops the antiproton comparison; the shared material is not an independent analysis or dataset.

\section{Phase orientation and the finite-history picture}
\label{sec:framework}

Particle and antiparticle modes appear in the standard field expansion with phases $e^{-ip\cdot x}$ and $e^{+ip\cdot x}$, respectively, with $p^0>0$. On these modes, the eigenvalue $K^\mu$ of the operator $i\partial^\mu$ is $+p^\mu$ or $-p^\mu$. The orientation convention is \cite{YangTime}
\begin{equation}
s_T=\operatorname{sign}(K^0),\qquad p^\mu=s_TK^\mu,\qquad p^0>0.
\label{eq:orientation}
\end{equation}
Here $s_T$ labels phase orientation, not the sign of the measured energy. Both members of a produced pair generate outgoing tracks in increasing laboratory time. Local QED amplitudes and causal detector records retain their usual interpretation.

The macroscopic content of the finite-history picture is a reduced effective overlap of the opposite phase history with the environment defined by matter. I represent its leading spectral effect by a single rescaling and a nearly constant relative amplitude. This is the phenomenological prescription of the framework, rather than a consequence of changing the sign of a local energy or merely relabeling an identical path. The positron comparison below applies this prescription to the measured electron reference.

Let $B(x)$ denote a reference spectrum, where $x$ is energy. Let $\xi$ be a constant dimensionless horizontal factor, and $\mathcal A$ the dimensionless relative amplitude. The response is
\begin{equation}
B(x)\longrightarrow\mathcal A B(\xi x),\qquad 0<\xi\leq1,
\label{eq:response}
\end{equation}
with $\mathcal A$ approximately constant over the interval considered. In logarithmic notation this is $d\ln\mathcal A/d\ln x\simeq0$: an amplitude proportional to $x^\alpha$ would instead add the index $\alpha$. Thus the approximation means that the history weight changes the amount of flux without introducing an additional leading spectral power.

Define $x_*$ as the position of a specified feature of the reference, such as the maximum of a stated energy-weighted spectrum. The same feature appears at $x_*/\xi$ after rescaling. The lepton comparison below uses this displacement of a broad spectral feature.

\section{Positrons: a displaced spectral scale}
\label{sec:positron}
\subsection{Less accumulated exposure, a higher spectral scale}
\label{sec:exposure}

Electrons and positrons obey the same leading radiative energy-loss law. Writing $t_M$ for matter-clock time, a useful approximation is \cite{BlumenthalGould}
\begin{equation}
\frac{dE}{dt_M}=-\beta u E^2,\qquad
\frac{1}{E}-\frac{1}{E_{\rm inj}}=\beta\mathcal X.
\label{eq:loss}
\end{equation}
Here $E_{\rm inj}$ is initial energy, $u$ the effective radiation and magnetic energy density, $\beta>0$ the loss coefficient, and $\mathcal X=\int u\,dt_M$ the accumulated exposure along the effective history. The inverse-Compton part uses the Thomson-limit scaling. The physical energy loss has the same sign for either charge; the effect considered is a difference in exposure, not positron acceleration.

In the present framework, positrons have a smaller effective overlap with the matter-defined environment. Labeling the two exposures by charge, $\mathcal X_+$ for positrons and $\mathcal X_-$ for electrons, let $q_x,q_y,q_z$ denote the corresponding effective overlap-reach ratios in the three spatial directions. The factorized exposure ratio is
\begin{equation}
\xi_G\equiv\frac{\mathcal X_+}{\mathcal X_-}
\simeq q_xq_yq_z\equiv q^3.
\label{eq:overlap}
\end{equation}
Only the product enters the comparison. The shorthand $q$ is its geometric mean, so equal directional reaches are not required. The illustrative choice $q=0.30$ gives $\xi_G=0.027$. These are effective history parameters, not measured ratios of Galactic spatial sizes.

Equation~\eqref{eq:loss} identifies the characteristic radiative-loss scale $(\beta\mathcal X)^{-1}$: reducing the exposure raises this scale. The single-scale description in Eq.~\eqref{eq:response} carries this inverse-exposure scaling to the corresponding lepton spectrum. The low-energy electron data will be used as an empirical reference for that response. Since these are top-of-instrument fluxes, they retain heliospheric effects \cite{AMSelectron}; their use here is a template comparison, not an extraction of an intrinsic interstellar cooling scale.

\subsection{A direct comparison without a spectral fit}
\label{sec:lepton-data}

To display the broad structure, define $\mathcal S_i(E)=E^3\Phi_i(E)$, with $\Phi_i$ the differential intensity of species $i$. Take $\mathcal S_{e,\rm ref}$ directly from the 22 tabulated electron points with representative energies between 4 and 20 GeV \cite{AMSelectron}. Line segments join successive values on a logarithmic energy axis; no log-parabola or other fitted spectral function is used. This preserves the measured local shape. To make both sides of the broad structure visible, dashed continuations use the adjacent electron data selected from 2 to 40 GeV, a factor-of-two extension of each selection limit. The central 4--20 GeV comparison is solid; no scale or amplitude is adjusted for the continuation.

\begin{figure*}[t]
\centering
\includegraphics[width=0.96\textwidth]{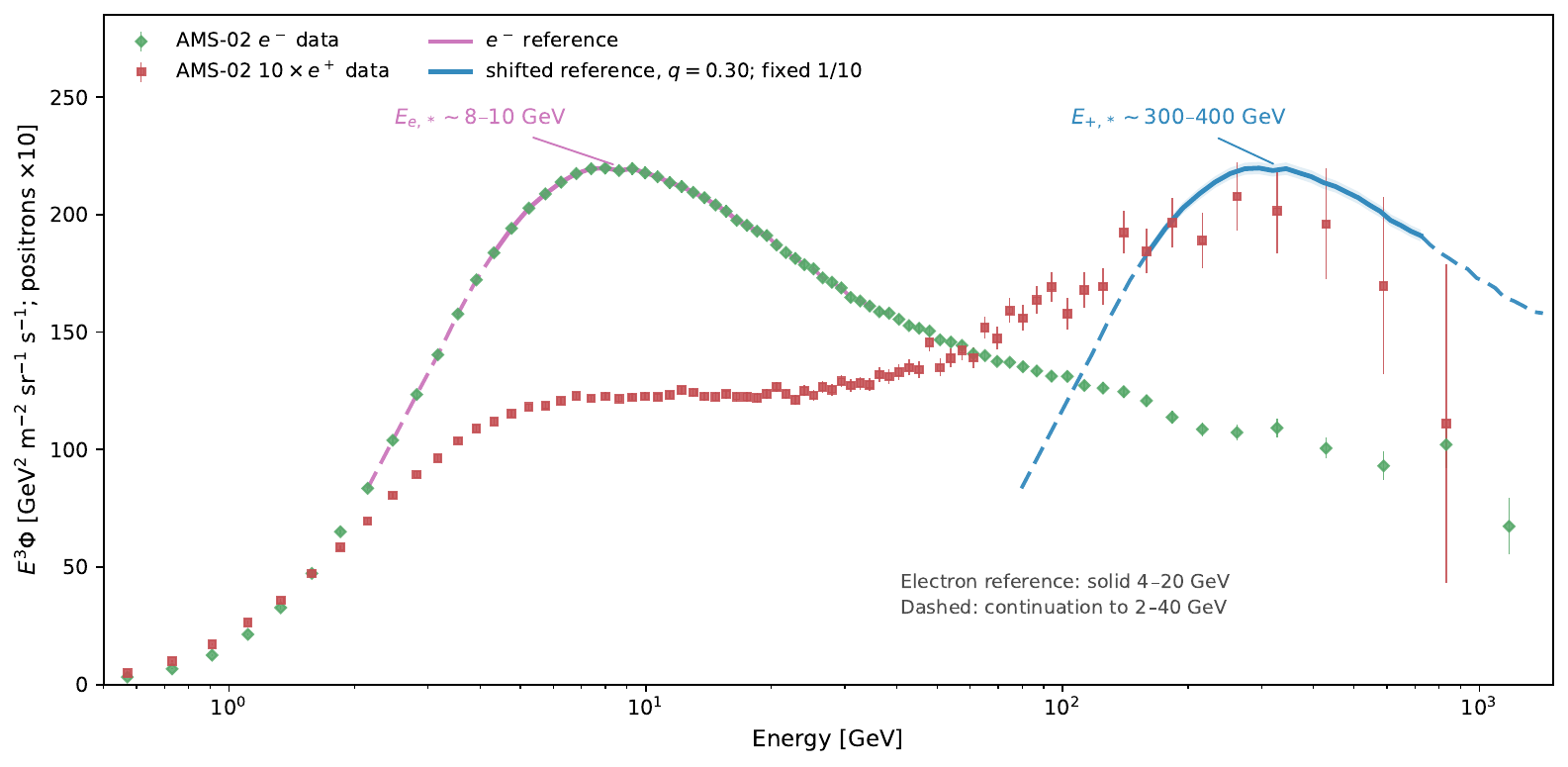}
\caption{AMS-02 electron data and positron data multiplied by ten, displayed as $E^3\Phi$ on a logarithmic energy axis \cite{AMSpositron,AMSelectron}. Purple curves join measured electron reference values; blue curves apply the fixed mapping with $q=0.30$ and relative weighted amplitude $1/10$ of Eq.~\eqref{eq:positron-map}. Solid segments show the central 22-point, 4--20 GeV selection (mapped representative energies 160--723 GeV). Dashed continuations use the adjacent measured electron entries in a 2--40 GeV selection, extending the mapped curve to about 80--1420 GeV without an analytic extrapolation or fit. They explore the mapping beyond the central comparison interval, rather than establish its validity there; the upper continuation extends beyond the 1 TeV upper edge of the positron data. The pale blue band on the solid segment carries the reference's vertical errors, not model uncertainty. Labels indicate broad maximum regions.}
\label{fig:positron}
\end{figure*}

For the qualitative comparison, the positron weighted spectrum is taken to be about one tenth of the reference at the corresponding, rescaled energy. The reason this amplitude scale is compatible with a number-conserving dilution picture is given in Sec.~\ref{sec:normalization}. The fixed prescription is
\begin{equation}
\begin{aligned}
\mathcal S_+(E)&\simeq\frac{1}{10}\mathcal S_{e,\rm ref}(\xi_GE),\\
\Phi_+(E)&\simeq\frac{\xi_G^3}{10}\Phi_{e,\rm ref}(\xi_GE).
\end{aligned}
\label{eq:positron-map}
\end{equation}
The second line makes the differential-flux normalization explicit: $\xi_G^3/10\simeq2\times10^{-6}$. The factor ten compares weighted spectra at different energies, not the positron and electron fluxes at the same energy. In Fig.~\ref{fig:positron}, multiplying the positron data and the mapped response by ten removes this fixed vertical factor visually. No positron amplitude is optimized.

The asterisk in $E_{e,*}$ denotes the location of the broad maximum of $\mathcal S_{e,\rm ref}$, near 8--10 GeV; $E_{+,*}$ denotes the corresponding maximum of $\mathcal S_+$. The broad scale therefore moves to
\begin{equation}
E_{+,*}=\frac{E_{e,*}}{\xi_G}\sim300\text{--}400\GeV.
\label{eq:peak}
\end{equation}
The ranges identify broad spectral regions, not fitted confidence intervals. The mapped position lies in the energy region where AMS observes the positron dropoff \cite{AMSpositron}; the maximum of $E^3\Phi$ and a fitted break or exponential cutoff have different definitions.

All published electron and positron points are retained for context. The full displayed reference contains 41 entries with representative energies 2.15--38.33 GeV, mapped to about 80--1420 GeV; 22 entries define the central comparison. Statistical and total systematic vertical errors are added in quadrature for display. The solid comparison retains the displaced broad structure and rounded amplitude without an added dominant source component. The dashed continuation falls below the positron data toward 80--120 GeV, while its high-energy decline is shallower than the trend of the highest positron central values. It is therefore not a description of the complete positron flux or a determination of the cutoff. In particular, the wider electron template is still empirical: its low-energy part retains heliospheric effects, and no separation of its spectral components is attempted.

\subsection{Why an amplitude of this order is possible}
\label{sec:normalization}

The geometric idea is that the same particle content, distributed over a larger effective region, gives a smaller local density. For similarly shaped, normalized three-dimensional profiles broadened by a linear factor $a$, the central density scales as $a^{-3}$ while the total particle number is unchanged. At an observer in the nearly uniform central region, the same factor multiplies the intensity for nearly isotropic ultrarelativistic particles, since intensity is proportional to local density \cite{PDGCosmic}.

To combine this dilution with the energy rescaling, take two finite electron and positron ensembles with the same injected differential count $\nu_{\rm inj}(E_{\rm inj})\propto E_{\rm inj}^{-\Gamma}$, including equal normalization over a common finite energy interval. Here $\Gamma$ is the injection index and $\nu_s(E)=dN_s/dE$ is the count spectrum after exposure, with $s=+,-$ labeling charge. For fixed $\mathcal X_s$, Eq.~\eqref{eq:loss} gives $E_{\rm inj}=E/(1-\beta\mathcal X_sE)$. Conservation in corresponding energy bins, $\nu_s(E)\,dE=\nu_{\rm inj}(E_{\rm inj})\,dE_{\rm inj}$, yields
\[
\begin{aligned}
\nu_s(E)&=\nu_{\rm inj}\!\left(\frac{E}{1-\beta\mathcal X_sE}\right)
          \frac{1}{(1-\beta\mathcal X_sE)^2}\\
&\propto E^{-\Gamma}(1-\beta\mathcal X_sE)^{\Gamma-2}.
\end{aligned}
\]
Since $\mathcal X_+=\xi_G\mathcal X_-$, the last factor is identical for $\nu_+(E)$ and $\nu_-(\xi_GE)$, giving $\nu_+(E)=\xi_G^\Gamma\nu_-(\xi_GE)$. Including the local spatial dilution therefore gives
\begin{equation}
\Phi_+(E)\simeq a^{-3}\xi_G^\Gamma\Phi_-(\xi_GE).
\label{eq:dilution}
\end{equation}
The inverse-map denominators must be positive, and both corresponding initial energies must lie in the common injection interval. This relation compares distributions at different initial energies, not a one-to-one rescaling of particles injected at the same energy. Each ensemble conserves its count over its own complete cooled interval. Here $\Phi_\pm$ denote the spectra of this illustration, not an independent derivation of the measured electron template. The factor $\xi_G^\Gamma$ arises from the common injected spectrum and the energy-bin Jacobian, not from the plotting weight $E^3$.

For representative values $\Gamma=2.5$ and $a=4$, the coefficient is $a^{-3}\xi_G^\Gamma\simeq1.9\times10^{-6}$. Its weighted-spectrum counterpart is $a^{-3}\xi_G^{\Gamma-3}\simeq0.095$, which motivates the rounded $1/10$ in Eq.~\eqref{eq:positron-map}. At fixed $\Gamma=2.5$, changing $a$ from 3 to 5 spans about $10^{-6}$--$4\times10^{-6}$ in the differential-flux coefficient. Thus a factor-of-a-few linear broadening can accommodate the amplitude scale in this example.

These illustrative choices were informed by the already-known relative spectral heights. They are an order-of-magnitude interpretation, not a blind normalization prediction. In particular, $a$ measures the width of a spatial distribution, whereas $q$ characterizes effective history overlap; a relation fixing one from the other has not been established. The qualitative result is that the required local dilution can be expressed without altering the common injected particle count.

\balance
\section{Discussion and conclusion}
\label{sec:discussion}

The finite-history picture separates the local interaction law from the accumulated overlap with the matter-defined environment. Both charges lose physical energy through the same radiative processes, while the positron response is assigned a smaller effective exposure. Represented by a single rescaling, that difference shifts the broad structure of an empirical electron reference into the high-energy positron region without changing the sign of the measured energy.

With $q=0.30$, the electron maximum region near 8--10 GeV maps to roughly 300--400 GeV. A weighted amplitude of $1/10$ is consistent in order of magnitude with the number-conserving spatial-dilution illustration. These are informed benchmarks, not independently measured Galactic parameters. The electron reference retains heliospheric effects, and the dashed continuations display differences from the positron data outside the central comparison. The construction therefore identifies a spectral-scale relation rather than the complete positron flux or its cutoff.

Within this scope, the high-energy positron pattern can be organized without introducing a special dominant astrophysical source to create the feature. Known populations, including pulsars, may contribute subleading components in this interpretation, but their fractions are not measured here. The empirical electron reference is a template for the common response, not a separation of all primary-electron and pair components. Establishing the history overlap and spatial dilution from a microscopic construction, and testing shape preservation over a broader energy range, remain the central physical questions. The present analysis isolates these questions without a detailed Galactic propagation model.

\section*{Declaration of competing interest}
The author declares that he has no known competing financial interests or personal relationships that could have appeared to influence the work reported in this paper.

\section*{Data availability}
The input spectra are the published AMS-02 electron and positron supplemental tables \cite{AMSpositron,AMSelectron}. The source package contains these two tables, code reproducing the fixed-scale figure, and numerical checks of the energy mapping, particle-count conservation, and normalized spatial dilution. No spectral or amplitude optimization is used for Fig.~\ref{fig:positron}.

\section*{Acknowledgments}
This work was supported by Academia Sinica, the National Science and Technology Council of Taiwan, and National Cheng Kung University.

\section*{Declaration of generative AI and AI-assisted technologies in the manuscript preparation process}
During preparation, the author used ChatGPT (OpenAI) for language editing, manuscript preparation, and analysis-code assistance. The author reviewed the output and takes full responsibility for the publication.


\begingroup
\small
\begin{thebibliography}{99}

\bibitem{AMSpositron}
M.~Aguilar \textit{et al.} (AMS Collaboration),
Towards Understanding the Origin of Cosmic-Ray Positrons,
Phys. Rev. Lett. \textbf{122} (2019) 041102.

\bibitem{AMSelectron}
M.~Aguilar \textit{et al.} (AMS Collaboration),
Towards Understanding the Origin of Cosmic-Ray Electrons,
Phys. Rev. Lett. \textbf{122} (2019) 101101.

\bibitem{Stueckelberg}
E.~C.~G. Stueckelberg,
La signification du temps propre en m\'ecanique ondulatoire,
Helv. Phys. Acta \textbf{14} (1941) 322.

\bibitem{Feynman}
R.~P. Feynman,
The Theory of Positrons,
Phys. Rev. \textbf{76} (1949) 749.

\bibitem{Peskin}
M.~E. Peskin and D.~V. Schroeder,
\textit{An Introduction to Quantum Field Theory}
(Addison-Wesley, 1995).

\bibitem{YangTime}
Y.~Yang,
Time Orientation in Quantum Field Theory,
Zenodo (2026),
\href{https://doi.org/10.5281/zenodo.20821313}{doi:10.5281/zenodo.20821313}.

\bibitem{YangYeung}
Y.~Yang and W.~B. Yeung,
Highly Suppressed Detection Probability of the Primordial Antimatter in the Present-Day Universe,
Chin. J. Phys. \textbf{77} (2022) 1544; arXiv:2604.02388.

\bibitem{HooperPulsars}
D.~Hooper, P.~Blasi, and P.~D. Serpico,
Pulsars as the Sources of High Energy Cosmic Ray Positrons,
JCAP \textbf{01} (2009) 025.

\bibitem{YangCombined}
Y.~Yang,
A Common Antimatter Response in AMS-02 Positrons and Antiprotons,
\href{https://arxiv.org/abs/2606.04043}{arXiv:2606.04043}.

\bibitem{BlumenthalGould}
G.~R. Blumenthal and R.~J. Gould,
Bremsstrahlung, Synchrotron Radiation, and Compton Scattering of High-Energy Electrons Traversing Dilute Gases,
Rev. Mod. Phys. \textbf{42} (1970) 237.

\bibitem{PDGCosmic}
J.~Alvarez-Mu\~niz, Z.~Cao, U.~F.~Katz, P.~Mertsch, and C.~Spiering,
Cosmic Rays, in \textit{Review of Particle Physics}, Particle Data Group,
revised March 2026,
\href{https://pdg.lbl.gov/2026/reviews/rpp2026-rev-cosmic-rays.pdf}{online review}.

\end{thebibliography}
\endgroup
\end{document}